\documentclass[10pt]{article}

\usepackage{amstext,amsmath,amssymb,amsfonts,bbm,slashed}
\usepackage[latin1]{inputenc}
\usepackage{epsfig}
\usepackage{hyperref}
\usepackage{amsthm}
\usepackage{subfigure}
\usepackage{color}
\usepackage{multirow}

\newtheorem{proposition}{Proposition}

\newcommand{\bea}{\begin{eqnarray}}
\newcommand{\eea}{\end{eqnarray}}
\newcommand{\beq}{\begin{equation}}
\newcommand{\eeq}{\end{equation}}

\newcommand{\bpro}{\begin{pro}}
	\newcommand{\epro}{\end{pro}}
\newcommand{\blem}{\begin{lem}}
	\newcommand{\elem}{\end{lem}}
\newcommand{\bdfn}{\begin{dfn}}
	\newcommand{\edfn}{\end{dfn}}
\newcommand{\bcor}{\begin{cor}}
	\newcommand{\ecor}{\end{cor}}
\newcommand{\bthm}{\begin{thm}}
	\newcommand{\ethm}{\end{thm}}
\newcommand{\bex}{\begin{ex}}
	\newcommand{\eex}{\end{ex}}
\newcommand{\brmk}{\begin{rmk}}
	\newcommand{\ermk}{\end{rmk}}
\newcommand{\bpr}{\begin{pr}}
	\newcommand{\epr}{\end{pr}}

\makeatletter

\begin{document}

	\begin{center}
		
		{\LARGE\bf   Noncommutative Kepler Dynamics:  symmetry groups and bi-Hamiltonian structures }

		\vspace{15pt}
		
		{\large   Mahouton Norbert Hounkonnou $^{\dagger} $ , Mahougnon Justin Landalidji $^{\dagger}, $  Melanija Mitrovi\'c $^{\ddagger} $
		}
		
		\vspace{15pt}
		
		{  $\dagger$)\,\,
			International Chair of Mathematical Physics
			and Applications  (ICMPA-UNESCO Chair)\\
			University of Abomey-Calavi, 072 B.P. 50 Cotonou, Republic of Benin\\
			E-mails: { norbert.hounkonnou@cipma.uac.bj with copy to
				hounkonnou@yahoo.fr} \\
			E-mails: { landalidjijustin@yahoo.fr}   \\
			\vspace{5pt}
			$\ddagger$)\,\,  Faculty of Mechanical Engineering, Department of Mathematics and Informatics\\
			University of Ni\v s,  Serbia \\
			E-mails: {  melanija.mitrovic@masfak.ni.ac.rs  }
		}

	\end{center}
	\vspace{10pt}
	\begin{abstract}
		Integrals of motion are constructed from noncommutative (NC) Kepler dynamics, generating $SO(3),$ $SO(4),$ and  $SO(1,3)$ dynamical symmetry groups. The Hamiltonian vector field is derived in action-angle coordinates, and the existence of a hierarchy of bi-Hamiltonian structures is highlighted. Then, a family of Nijenhuis recursion operators is computed and discussed.
		
		
	\end{abstract}
	
	\section{Introduction}\label{sec1}
	
	In {\it Mysterium Cosmographicum}, published in 1596,  Kepler
	proposed a model of the solar system by relating the five
	extra-terrestrial planets (Mercury, Venus,  Mars, Jupiter, and
	Saturn) known at that time to be the five Platonic solids (the
	tetrahedron or pyramid, cube, octahedron, dodecahedron, and
	icosahedron) \cite{liv,zho}. Kepler's work attracted the attention
	of the Danish astronomer Brahe who recruited  him in October 1600 as
	an assistant. Kepler then accessed  Brahe's
	empirical data, and published the so-called Kepler's first two laws of planetary motion  in 1609 \cite{Voe,kepl1}, and the third one in 1619 \cite{kepl2}.
	Kepler thus obtained from Brahe a detailed set of observations of the motion
	of the planet Mars,  analyzed them and deduced that the path of
	Mars is an ellipse, with the sun located at one of its  focal points, and that
	the radius vector from the sun to this planet sweeps out equal areas in
	equal times \cite{Brac}.  The Kepler direct problem of determining  the nature of the
	force required to maintain elliptical motion about a focal force center  was finally solved by Newton in the 1680s.
	Indeed, Newton determined the functional dependence on distance of the force required to sustain such an elliptical path
	of Mars about the sun as a center of force located at a focal point of the ellipse. Today scientists still concentrate on
	the inverse problem that consists in using the combined gravitational forces of the sun and the other
	planets to predict and explain perturbations in the conic paths of planets and comets.
	
	On the other hand, although the Kepler problem was a central theme of analytical dynamics for centuries,
	addressed by several authors, it  continues to be so in the contemporary studies as well, revealing interesting mathematical symmetries.
	Significant steps for understanding the symmetries underlying the Kepler dynamics were made  with its  quantum explorations, where the
	$SO(4)$,  $O(4,1),$ and $O(4,2)$ symmetry groups were explored (see $e.g.,$   \cite{len,pa,foc,bar,ban1,ban2,hul} and the references therein).
	These studies led to the re-investigation of the classical Kepler problem (see \cite{fra,bac,gior,gui} and the references therein).
	Thus, in 1966,  Bacry {\it et al.}  \cite{bac} proved that the transformations generated by the angular momentum and the Runge-Lenz vector indeed form a group
	of canonical transformations isomorphic to $SO(4).$
	In 1968,  Gy\"{o}rgyi \cite{gior} gave a formulation of the Kepler problem, manifestly invariant with respect to the $SO(4)$ and $SO(3,1)$ symmetry groups,
	in terms of the Fock variables and their canonical conjugates,  respectively, and introduced a new time parameter, proportional to the eccentric anomaly.
	In that work, a transformation of the dynamical variables was performed in order to regain the standard time $t$ leading in a natural way to Bacry's generators
	inducing an $SO(4,2)$  symmetry group. In 1970, Moser \cite{mos}  regularized the Kepler problem,   enlarging the phase space in such a way that
	the temporal evolution generates a global time flow, a situation
	otherwise precluded by the existence of collision orbits. Six years
	later, $i.e.,$ in 1976,  Ligon {\it et al.} \cite{lig}  completed
	the previous works by adding the symplectic forms and the
	Hamiltonian vector fields.  They transformed the Kepler problem in
	such a way that both the time flow and the $SO(4)$ symmetry were
	globally realized in a simple and canonical way. This was also done
	for a positive energy and the group $SO(1,3).$ Since 1976, several
	works have focused on the classical  Kepler problem, (see $e.g.,$
	\cite{chan,chen,cuch,gui,mil,marl}). Further, in a remarkable book
	published in 2001,  Vilasi \cite{vil2} showed that the Lie algebra
	of symmetries for the Kepler dynamics is twice $so(3),$ or better,
	$su(2) \otimes su(2),$ which is locally isomorphic with  $so(4).$
	
	In addition, in the last few decades,  there was a renewed interest in
	the Kepler problem as one of completely integrable Hamiltonian systems
	(IHS), the concept of which goes back to Liouville in 1897 \cite{lio}
	and Poincar\'{e} in 1899 \cite{pau}.  Loosely speaking, IHS are
	dynamical systems admitting a Hamiltonian description, and possessing
	sufficiently many constants of motion. Many of these systems are
	Hamiltonian systems with respect to two compatible symplectic
	structures \cite{mag1,gel,vil1,fil1} permitting
	a geometrical interpretation of the so-called recursion operator
	\cite{lax}. Hence,  a natural approach to integrability is to try to
	find sufficient conditions for the eigenvalues of the recursion
	operator to be in involution \cite{san}. In 1992, Marmo {\it et al.}
	\cite{mar} constructed two Hamiltonian structures for the Kepler
	problem in angle-action coordinates  and proved their compatibility
	condition by checking the vanishing of the
	Nijenhuis tensor of the corresponding recursion operator.
	
	Over the past few years, Magri's approach  \cite{mag1} to
	integrability through bi-Hamiltonian structures has became one of
	the most powerful methods relating to the integrability of evolution
	equations, applicable in studying both finite and infinite
	dimensional dynamical systems \cite{gri}. This approach has also
	been proven to be one of  the classical methods of integrability of
	evolution equations along with, for example, the Hamilton-Jacobi
	method of separation of variables and the method of the Lax
	representation \cite{lax,smir}.
	When a completely
	integrable Hamiltonian system does  admit a bi-Hamiltonian construction,
	one can generate infinite hierarchies of conserved quantities through the construction by Oevel \cite{oev} based on scaling invariances and master
	symmetries \cite{fer2,smir2}.
	Another generalization is due to  Bogoyavlenskij \cite{bog} who
	proposed a complete classification of the invariant Poisson
	structures for non-degenerate and degenerate Hamiltonian systems. In
	1997 and 1999, Smirnov \cite{smir,smir2} formulated a constructive
	method of transforming a completely integrable Hamiltonian system,
	in Liouville's sense, into Magri-Morosi-Gel'fand-Dorfman's (MMGD)
	bi-Hamiltonian form. He showed that the action-angle variables can
	be a powerful tool in solving the problem of transformation of a
	completely integrable Hamiltonian system into its MMGD
	bi-Hamiltonian form. The application of Smirnov's result to the
	classical Kepler problem is due to the possibility of transforming
	the action-angle coordinates  connected with the spherical-polar
	coordinates to the Delaunay coordinates. 
	In 2005,  Ra\~{n}ada  \cite{ra}
	proved the existence of a bi-Hamiltonian structure arising from a
	non-symplectic symmetry  as well as the existence of master
	symmetries and additional integrals of motion (weak
	superintegrability) for certain particular values of two parameters
	$b$ and $k$. In 2015, Grigoryev {\it et al.} showed that the
	perturbed Kepler problem is a bi-Hamiltonian system in spite of the
	fact that the graph of the Hamilton function is not a hypersurface
	of translation, which goes against a necessary condition for the
	existence of the bi-Hamiltonian structure according to the Fernandes
	theorem \cite{fern,gri}.  They explicitly presented a few
	non-degenerate bi-Hamiltonian formulations of the perturbed Kepler
	problem using the Bogoyavlenskij construction of a continuum of
	compatible Poisson structures for the isochronous Hamiltonian
	systems \cite{bog}.   Our present work focuses on an investigations
	into dynamical symmetry groups and bi-Hamiltonian structures for the
	Kepler dynamics in a noncommutative (NC) phase space.
	
	The paper is organized as follows.  In Section \ref{sec2}, we
	present the considered noncommutative phase space and give some
	basic notions useful for our subsequenst development. In Section
	\ref{sec3}, we give the
	Hamiltonian function, symplectic form and vector field describing
	the Kepler dynamics  in the noncommutative phase space. In Section \ref{sec4}, we study the existence
	of  dynamical symmetry groups $SO(3),$ $SO(4),$ and $SO(1,3)$  in the
	described setting. In Section \ref{sec5}, we derive relevant
	geometric quantities in action-angle variables, and obtain the
	corresponding Hamiltonian system. In Section \ref{sec6}, we
	construct bi-Hamiltonian structures and  the associated recursion
	operators. In Section \ref{sec7}, we define the hierarchy of master
	symmetries and compute the conserved quantities. In Section
	\ref{sec8}, we end with some concluding remarks.
	
	\section{Noncommutative phase space and basic definitions} \label{sec2}
	Let 
	$\mathcal{Q} = \mathbb{R}^{3} \backslash \{0\}$ be  the manifold  describing the configuration space of the Kepler problem,
	and $\mathcal{T}^{\ast}\mathcal{Q} = \mathcal{Q} \times \mathbb{R}^{3}$ be
	the cotangent bundle with the local coordinates $(q,p),$ and
	a natural symplectic
	structure $\omega :
	\mathcal{T}\mathcal{Q} \longrightarrow \mathcal{T}^{\ast}\mathcal{Q}$  given by
	{\small\begin{equation*}\label{Ksy}
		\omega = \sum_{i = 1}^{3} dp_{i} \wedge dq^{i},
		\end{equation*} }
	where $\mathcal{T}\mathcal{Q}$
	is the tangent bundle.
	By definition, $\omega$ is
	non-degenerate. It induces the map $P$:
	$\mathcal{T}^{\ast}\mathcal{Q} \longrightarrow \mathcal{T}\mathcal{Q}$
	defined by
	{\small\begin{equation*}\label{Kbi}
		P = \sum_{i = 1}^{3} \dfrac{\partial}{\partial p_{i} }\wedge
		\dfrac{\partial}{\partial q^{i}},
		\end{equation*}}  called a bivector field, which is  the inverse map of $\omega, $ i.e.   $ \omega \circ P = P \circ \omega = 1$
	\cite{vil2}. In this case, the Hamiltonian vector field $X_{f}$
	of a Hamiltonian function $f$ is given by
	{\small\begin{equation*}\label{Kvec}
		X_{f} = Pdf.
		\end{equation*} }
	The noncommutativity \cite{alinc2} between phase space variables is here understood by replacing the usual product with the $\beta-$star product, also known as the Moyal product, between two arbitrary functions of position and momentum as follows
	\cite{vak,ma,khos1}:
	{\small\begin{equation*} \label{Eq_3_1}
		(f\ast_{\beta}g) (q,p) = f(q_{i},p_{i}) \exp{\bigg(\dfrac{1}{2}\beta^{ab}\overleftarrow{\partial}_{a}\overrightarrow{\partial_{b}}\bigg)} g(q_{j},p_{j})\Bigg|_{(q_{i},p_{i})= (q_{j},p_{j})},
		\end{equation*}}
	where
	{\small\begin{equation*} \label{Eq_3_2}
		\beta_{ab} = \left(
		\begin{array}{cc}
		\alpha_{ij} & \delta_{ij} + \gamma_{ij} \\
		- \delta_{ij} - \gamma_{ij} & \lambda_{ij} \\
		\end{array}
		\right).
		\end{equation*}}
	The parameters
	$\alpha$ and $\lambda$ are antisymmetric  $n \times n$ matrices generating the noncommutativity in coordinates and momenta, respectively;   $\gamma$ can depend on $\alpha$ and $\lambda.$
	The     $\ast_{\beta}$ deformed Poisson bracket is defined as:
	{\small \begin{equation*} \label{Eq_3_3}
		\{f,g\}_{\beta} := f\ast_{\beta}g - g\ast_{\beta}f,
		\end{equation*}}
	providing the commutation relations
	{\small\begin{equation}\label{Eq_3_4}
		\{q_{i},q_{j}\}_{\beta} = \alpha_{ij}, \ \{q_{i},p_{j}\}_{\beta} =\delta_{ij} + \gamma_{ij}, \ \{p_{i},q_{j}\}_{\beta} = - \delta_{ij} - \gamma_{ij}, \ \{p_{i},p_{j}\}_{\beta} =\lambda_{ij}.
		\end{equation}}
	It is worth noticing that the following transformed coordinates:
	{\small\begin{equation}\label{Eq_3_5}
		q'_{i} = q_{i} - \dfrac{1}{2}\sum_{j=1}^{n}\alpha_{ij}p_{j}, \quad p'_{i} = p_{i} + \dfrac{1}{2}\sum_{j=1}^{n}\lambda_{ij}q_{j},
		\end{equation}}
	obey the same commutation relations as in  \eqref{Eq_3_4} with respect to the usual Poisson bracket, i.e.
	{\small\begin{equation*} \label{Eq_3_6}
		\{q'_{i},q'_{j}\} = \alpha_{ij}, \ \{q'_{i},p'_{j}\}=\delta_{ij} + \gamma_{ij}, \ \{p'_{i},q'_{j}\}= -\delta_{ij} - \gamma_{ij}, \ \{p'_{i},p'_{j}\} =\lambda_{ij},
		\end{equation*}}
	while $q_{i}$ and $p_{j}$ satisfy the  canonical commutation relations:
	{\small\begin{equation*} \label{Eq_3_7}
		\{q_{i},q_{j}\} = 0, \quad \{q_{i},p_{j}\}=\delta_{ij}, \quad \{p_{i},p_{j}\}=0.
		\end{equation*}}

	A Hamiltonian system is a triple $(\mathcal{Q}, \omega , H )$, where $(\mathcal{Q}, \omega)$ is a symplectic manifold, which is,  in the present context, the configuration  space for the Kepler problem, and $H$  is a smooth function
	on $\mathcal{Q}$, called  {\it Hamiltonian} or  {\it Hamiltonian function} \cite{rud}.

	Given a general dynamical system defined on the $2n$-dimensional manifold $\mathcal{Q}$ \cite{smir}, its evolution can be described by the equation
	{\small\begin{equation} \label{dyn}
		\dot{x}(t) = X(x), \quad x \in \mathcal{Q}, \quad X \in \mathcal{T}\mathcal{Q}.
		\end{equation}}
	If the system \eqref{dyn} admits two different Hamiltonian representations:
	{\small\begin{equation*} \label{dyn2}
		\dot{x}(t) = X_{H_{1},H_{2}} = P_{1}dH_{1} = P_{2}dH_{2},
		\end{equation*}}
	its integrability as well as many other properties are subject to Magri's approach. The bi-Hamiltonian vector field $X_{H_{1},H_{2}}$ is defined by two pairs of Poisson bivectors $P_{1}, P_{2}$ and Hamiltonian functions $ H_{1},H_{2}.$ Such a  manifold $\mathcal{Q}$ equipped with two Poisson bivectors is called a double Poisson manifold, and the quadruple $(\mathcal{Q},P_{1},P_{2},X_{H_{1},H_{2}} )$ is  called a bi-Hamiltonian system.   $P_{1}$ and $P_{2}$ are  two compatible Poisson bivectors with vanishing Schouten-Nijenhuis bracket \cite{du}: 
	$$[P_{1}, P_{2}]_{NS} = 0.$$
	
	\section{NC Kepler  Hamiltonian system}\label{sec3}
	
	We consider the NC Kepler Hamiltonian function in the transformed phase space coordinates (\ref{Eq_3_5}),
	{\small\begin{align*}
		H' = \sum_{i=1}^{3} \dfrac{p'_{i}p'^{i}}{2m} - \dfrac{k}{r'}.
		\end{align*}}
	The dynamical variables
	satisfy the  NC Poisson bracket \cite{hkn1}
	{\small\begin{equation*} \label{Eq_3_10}
		\{f,g\}_{nc} = \sum_{\nu=1}^{3}\theta_{\nu}^{-1}\Bigg(\dfrac{\partial{f}}{\partial{p_{\nu}}}\dfrac{\partial{g}}{\partial{q^{\nu}}} - \dfrac{\partial{f}}{\partial{q^{\nu}}}\dfrac{\partial{g}}{\partial{p_{\nu}}}\Bigg)
		\end{equation*}}
	with respect to the
	NC symplectic form
	{\small\begin{equation*} \label{Eq_form}
		\omega_{nc} := \displaystyle\sum_{i=1}^{3}dp'_{i}\wedge dq'^{i} =  \sum_{\nu=1}^{3}\theta_{\nu}dp_{\nu}\wedge dq^{\nu},        \end{equation*}}    with
	{\small \begin{equation*}\label{sys2}   \theta_{\nu} = \displaystyle\sum_{\mu = 1}^{4}\Bigg(\delta_{\mu\nu} + \dfrac{1}{4}\lambda_{\mu\nu}\alpha_{\mu\nu}\Bigg) \neq 0, \ \ \delta_{\mu\nu} = \left\{
		\begin{array}{ll}
		0, & \mbox{if} \ \mu \neq \nu \\
		1, & \mbox{if} \ \mu = \nu.
		\end{array}
		\right.
		\end{equation*}    }
	For $n=3,$ the Hamiltonian function $H'$ takes the form
	{\small\begin{align}\label{Ham}
		H' = \dfrac{1}{2m}\sum_{i=1}^{3} \bigg(p_{i} + \dfrac{1}{2}\sum_{j=1}^{3}\lambda_{ij}q^{j}\bigg)^{2} -k \bigg[\sum_{i=1}^{3}\bigg(q^{i} - \dfrac{1}{2}\sum_{j=1}^{3}\alpha_{ij}p_{j} \bigg)^{2}\bigg]^{-1/2}
		\end{align}}
	yielding the Hamilton's equations:
	{\small\begin{eqnarray}
		\dot{q}^{\mu} &:=& \{ H', q^{\mu} \}_{nc}= \theta^{-1}_{\mu} \Bigg( \sigma_{\mu}p_{\mu} + \sum_{s=1}^{3} R_{\mu s}q^{s} + \dfrac{k}{4Y^{3}}\sum_{l,\nu=1}^{3}\alpha_{l\mu}\alpha_{l\nu}p_{\nu}\bigg)\label{Hamile1} \\
		\dot{p}_{\mu} &:=& \{ H', p_{\mu}\}_{nc}= - \theta^{-1}_{\mu}\Bigg( \tilde{\sigma}_{\mu}q^{\mu} - \sum_{s=1}^{3} R_{\mu s}p_{s} + \dfrac{1}{4m}\sum_{l,\nu=1}^{3}\lambda_{l\mu}\lambda_{l\nu}q^{\nu}  \Bigg), \label{Hamile2}
		\end{eqnarray}}
	where
	
	{\small\[Y = \bigg[\bigg(q^{1} - \dfrac{1}{2}(\alpha_{12}p_{2} + \alpha_{13}p_{3}) \bigg)^{2} + \bigg(q^{2} - \dfrac{1}{2}(\alpha_{21}p_{1} + \alpha_{23}p_{3}) \bigg)^{2} + \bigg(q^{3} - \dfrac{1}{2}(\alpha_{31}p_{1} + \alpha_{32}p_{2}) \bigg)^{2} \bigg]^{ 1/2}, \]
		\[  \sigma_{\mu} = \dfrac{1}{m} +  \dfrac{1}{4Y^{3}}\sum_{i=1}^{3} (\alpha_{i\mu})^{2},  \quad \tilde{\sigma}_{\mu} = \dfrac{1}{Y^{3}} +  \dfrac{1}{4m}\sum_{i=1}^{3} (\lambda_{i\mu})^{2}, \quad  \mbox{and} \quad  R_{\mu s} = \dfrac{\lambda_{\mu s}}{2m} - \dfrac{\alpha_{s\mu}k}{2Y^{3}}. \] }
	In terms of the new coordinates $p'_{i}$ and $q'^{i},$ the Hamilton's equations \eqref{Hamile1} and \eqref{Hamile2} become
	{\small \begin{align*}
		& \dot{q}^{i} = \sum_{j=1}^{3} \theta^{-1}_{i}  \Bigg(  \dfrac{1}{m}p'_{i} + \dfrac{k}{2Y^{3}}\alpha_{ij}q'^{j} \Bigg), \
		\dot{p}_{i} = \sum_{j=1}^{3} \theta^{-1}_{i}\Bigg(\dfrac{1}{2m}\lambda_{ji}p'_{j} + \dfrac{k}{Y^{3}}q'^{i}\Bigg),
		\end{align*}}
	where $\mu, i = 1,2,3,    \ \nu \neq\mu; $
	$q'^{i} = q'^{i} (q,p) $ and  $p'_{i} =p'_{i}( q,p)$ are given by the relations \eqref{Eq_3_5}.
	
	The Hamiltonian function $H'$ \eqref{Ham} can be viewed as a generalization of the Hamiltonian function obtained in our previous work \cite{hkn2}.
	The quantity $Y$  deforming the distance  beetween the sun and the considered planet is responsible for the distortion of the  conic path about the sun.
	
	The $1$-form  $dH' \in \mathcal{T}^{\ast}\mathcal{Q}$  is given by
	{\small\begin{align} \label{Eq_dh}
		dH' &= \sum_{\mu=1}^{3} \Bigg[ \Bigg( \sigma_{\mu}p_{\mu} + \sum_{s=1}^{3} R_{\mu s}  q^{s} + \dfrac{k}{4Y^{3}}\sum_{l,\nu=1}^{3}\alpha_{l\mu}\alpha_{l\nu}p_{\nu}  \Bigg)dp_{\mu} \cr
		&+ \Bigg( \tilde{\sigma}_{\mu}q^{\mu} - \sum_{s=1}^{3}  R_{\mu s}p_{s} + \dfrac{1}{4m}\sum_{l,\nu=1}^{3}\lambda_{l\mu}\lambda_{l\nu}q^{\nu}  \Bigg)dq^{\mu}\Bigg],
		\end{align}}
	where $\ \nu \neq\mu. $
	
	Using the  NC Poisson bracket \eqref{Eq_3_10}, we obtain the  following NC Hamiltonian vector field
	{\small\begin{align*}\label{Eq_vect}
		X_{H'}& = \sum_{\mu=1}^{3}  \theta^{-1}_{\mu} \Bigg[ \Bigg( \sigma_{\mu}p_{\mu} + \sum_{s=1}^{3} R_{\mu s}q^{s} + \dfrac{k}{4Y^{3}}\sum_{l,\nu=1}^{3}\alpha_{l\mu}\alpha_{l\nu}p_{\nu}  \Bigg)\dfrac{\partial}{\partial{q^{\mu}}} \cr
		& - \Bigg( \tilde{\sigma}_{\mu}q^{\mu} - \sum_{s=1}^{3} R_{\mu s}p_{s} + \dfrac{1}{4m}\sum_{l,\nu=1}^{3}\lambda_{l\mu}\lambda_{l\nu}q^{\nu}  \Bigg)\dfrac{\partial}{\partial{p_{\mu}}}\Bigg], \ \nu \neq\mu
		\end{align*}}
	satisfying the required condition for a Hamiltonian system, i. e.
	\begin{equation*}
	\iota_{_{X_{H'}}} \omega'= -dH',
	\end{equation*}
	where $\iota_{_{X_{H'}}} \omega'$ is  the interior product of $ \omega'$ with respect to the Hamiltonian vector fied $X_{H'}.$  Hence, the triplet$(\mathcal{T}^{\ast}\mathcal{Q},\omega',H')$ is a Hamiltonian system.
	
	The NC coordinates $q'^{i}$ and $ p'_{i}$ generate the following noncommutative relations:
	{\small\begin{equation*}
		\{ p'_{i}, q'^{j} \}_{nc} =  F_{ij}, \quad  \{ p'_{i}, p'_{j} \}_{nc} = D_{ij}, \quad \{ q'^{i}, q'^{j} \}_{nc} = E_{ij},
		\end{equation*}}
	where
	{\small \begin{align*}
		& F_{ii} = \theta_{i}^{-1} + \dfrac{1}{4}\sum_{j = 1}^{3}\lambda_{ij}\alpha_{ij} \theta_{j}^{-1}; \  F_{ij} = \dfrac{1}{4}\lambda_{ir}\alpha_{jr} \theta_{r}^{-1}; \  D_{ij} = \dfrac{\lambda_{ji}}{2}\bigg(\theta_{i}^{-1} + \theta_{j}^{-1} \bigg); \cr
		& \ E_{ij} = \dfrac{\alpha_{ji}}{2}\bigg(\theta_{i}^{-1} + \theta_{j}^{-1} \bigg), i,j,r = 1,2,3.
		\end{align*}}

	\section{Dynamical symmetry groups}\label{sec4}
	Let us start by defining
	the NC phase space angular momentum vector  $L'$ and the Laplace-Runge-Lenz (LRL) vector $A'$  as follows:
	{\small \begin{equation*}
		L' = q'\times p', \quad  A' = p' \times L' - mk\dfrac{q'}{Y},
		\end{equation*}}
	where $p'$ is the momentum vector, $q'$ is the position vector of the
	particle of mass $m.$ Their components
	{\small \begin{equation*}
		\begin{cases}
		L'_{1} = q'^{2}p'_{3} - q'^{3}p'_{2}
		\\
		L'_{2} = q'^{3}p'_{1} - q'^{1}p'_{3}
		\\
		L'_{3} = q'^{1}p'_{2} - q'^{2}p'_{1}
		\end{cases};
		\quad
		\begin{cases}
		A'_{1} = p'_{2}L'_{3} - p'_{3}L'_{2} - mk\dfrac{q'^{1}}{Y}
		\\
		\\
		A'_{2} = p'_{3}L'_{1} - p'_{1}L'_{3} - mk\dfrac{q'^{2}}{Y}
		\\
		\\
		A'_{3} = p'_{1}L'_{2} - p'_{2}L'_{1} - mk\dfrac{q'^{3}}{Y}
		\end{cases}
		\end{equation*}}
	do not commute with the Hamiltonian function $H',$ i.e.
	{\small\begin{eqnarray}
		\{ H', L'_{i} \}_{nc} &=&  \sum_{\mu,\nu,j = 1}^{3} \varepsilon_{\mu i \nu} \bigg[ \bigg(\dfrac{1}{m}D_{\nu j}p'_{j} + \dfrac{k}{Y^{3}}F_{\nu j}q'^{j} \bigg)q'^{\mu} + \bigg(\dfrac{1}{m}F_{j\nu}p'_{j} + \dfrac{k}{Y^{3}}E_{j\nu}q'^{j} \bigg)p'_{\mu} \bigg], \label{PB_HL_i} \\
		\{ H', A'_{i} \}_{nc}&=&  \sum_{\eta,\varrho = 1}^{3} \varepsilon_{i \eta \varrho} \bigg\{  \sum_{\mu,\nu,j = 1}^{3} \varepsilon_{\mu \varrho \nu} \bigg[ \bigg(\dfrac{1}{m}D_{\nu j}p'_{j} + \dfrac{k}{Y^{3}}F_{\nu j}q'^{j} \bigg)q'^{\mu} \nonumber\\
		&+& \bigg(\dfrac{1}{m}F_{j\nu}p'_{j} + \dfrac{k}{Y^{3}}E_{j\nu}q'^{j} \bigg)p'_{\mu} \bigg]p'_{\eta}
		+ \sum_{j = 1}^{3}\bigg(\dfrac{1}{m}D_{\eta j}p'_{j} + \dfrac{k}{Y^{3}}F_{\eta j}q'^{j} \bigg)L'_{\eta} \bigg\} \nonumber\\
		&-& \dfrac{mk}{Y}\sum_{j = 1}^{3}\bigg(\dfrac{1}{m}F_{ji}p'_{j} + \dfrac{k}{Y^{3}}E_{ji}q'^{j} \bigg)
		+ \dfrac{mk}{Y^{3}}\sum_{j,h = 1}^{3} \bigg(\dfrac{1}{m}F_{jh}p'_{j} + \dfrac{k}{Y^{3}}E_{jh}q'^{j} \bigg)q'^{h}q'^{i}, \nonumber\\  \label{PB_HA_i}
		\end{eqnarray} }
	where $ i = 1, 2, 3;$ $\varepsilon_{\mu \varrho \nu}, \varepsilon_{i \eta \varrho}, $  and $\varepsilon_{\mu i \nu}$  are Levi-Civita symbols giving by
	{\small \[
		\varepsilon_{i j k} := \dfrac{1}{2}(i-j)(j-k)(k-i)
		\]  }
	
	\begin{proposition} \label{Pro_3}
		\label{Pro_1} Under the following conditions:
		\begin{enumerate}
			\item [(1)]   $\lambda_{ij}\alpha_{ij} = - \bigg[\dfrac{1}{2}(\lambda_{i\kappa }\alpha_{i\kappa } + \lambda_{j\kappa }\alpha_{j\kappa }) + 4\bigg],   \quad i, j,  \kappa = 1, 2, 3,$
			\item [(2)] $\dfrac{q'^{i}}{q'^{j}} = \dfrac{\lambda_{\kappa i} \theta_{j}}{\lambda_{\kappa j} \theta_{i}} = - \dfrac{\alpha_{\kappa j} \theta_{i}}{\alpha_{\kappa i} \theta_{j}} \ \mbox{and}\ \dfrac{p'_{i}}{p'_{j}} =  \dfrac{\alpha_{\kappa i} \theta_{j}}{\alpha_{\kappa j} \theta_{i}} = - \dfrac{\lambda_{\kappa j} \theta_{i}}{\lambda_{\kappa i} \theta_{j}},$
			\item [(3)] $F$ be a symmetric  $3 \times 3$ matrices, where $ F_{11} = F_{22} = F_{33},$
		\end{enumerate}
		the vectors $L_i'$ and $A_i'$ are in involution with the Hamiltonian function $H',$  $i.e.,$
		{\small\begin{equation} \label{cnst}
			\{ H', L'_{i}\}_{nc} = 0, \ \{ H', A'_{i}\}_{nc} = 0,
			\end{equation}}
		and, hence,  become constants of motion or first integrals of   $H'$ on  $\mathcal{T}^{\ast}\mathcal{Q}.$    \end{proposition}
	\textbf{Proof.}
	
	The condition $(1)$ leads to
	$\lambda_{ij}\theta_{i} = \lambda_{ji}\theta_{j} \ \ \mbox{and} \ \ \alpha_{ij}\theta_{i} = \alpha_{ji}\theta_{j}.$ Therefore, $D_{ij} = E_{ij} = 0$, and the equations \eqref{PB_HL_i} and \eqref{PB_HA_i}  are reduced to
	{\small\begin{eqnarray*} \label{PB_HL_i_2}
			\{ H', L'_{i} \}_{nc} &=&  \sum_{\mu,\nu,j = 1}^{3} \varepsilon_{\mu i \nu} \bigg(  \dfrac{k}{Y^{3}}F_{\nu j}q'^{j} q'^{\mu} + \dfrac{1}{m}F_{j\nu}p'_{j}p'_{\mu} \bigg), \quad i = 1, 2, 3,\\
			\{ H', A'_{i} \}_{nc}&=&  \sum_{\eta,\varrho = 1}^{3} \varepsilon_{i \eta \varrho} \bigg\{  \sum_{\mu,\nu,j = 1}^{3} \varepsilon_{\mu \varrho \nu} \bigg(  \dfrac{k}{Y^{3}}F_{\nu j}q'^{j}q'^{\mu} + \dfrac{1}{m}F_{j\nu}p'_{j} p'_{\mu} \bigg)p'_{\eta} + \sum_{j = 1}^{3}  \dfrac{k}{Y^{3}}F_{\eta j}q'^{j}L'_{\eta} \bigg\} \label{PB_HA_i_2} \\
			&-& \dfrac{mk}{Y}\sum_{j = 1}^{3}\dfrac{1}{m}F_{ji}p'_{j} + \dfrac{mk}{Y^{3}}\sum_{j,h = 1}^{3}\dfrac{1}{m}F_{jh}p'_{j}q'^{h}q'^{i}, \quad i = 1,2,3.
	\end{eqnarray*}}
	After computing, replacing the $F_{ij}$ by their expressions,  and using the conditions $(2)$ and $(3)$, we obtain
	$\{ H', L'_{i} \}_{nc}= \{ H', A'_{i} \}_{nc}= 0.$
	Then, $L'$ and $A'$ are constants of motion or first integrals of $H'$ on $\mathcal{T}^{\ast}\mathcal{Q}.$
	
	$\hfill{\square}$
	
	Equation \eqref{cnst} means that the functions
	$L'_{i}$ and $A'_{i}, i = 1,2, 3,$ are also constant along the orbits of $H'$.
	Then, in $\mathcal{T}^{\ast}\mathcal{Q},$  the orbits of $H'$  lie in the inverse image of a value of these functions \cite{lig,gui}.
	
	The previous proposition naturally leads to the following:
	\begin{proposition}\label{Pro_2}
		Let  $L'$ be an integral of motion on $\mathcal{T}^{\ast}\mathcal{Q}.$ Then, under the defined NC Poisson bracket \eqref{Eq_3_10}, its components  $L'_{i}$'s  generate the Lie algebra $so(3)$ of the group $SO(3)$, $i.e., \  \{L_{i}' , L_{j}' \}_{nc} =  \varepsilon_{i j h} F'_{hh}L'_{h},$ where  $\varepsilon_{i j h} F'_{hh}$ are the structure constants of the Lie algebra.
	\end{proposition}
	\textbf{Proof.}
	
	Under the conditions $(1), (2),$ and $(3)$ of  Proposition \ref{Pro_1},  the Poisson bracket
	{\small\begin{eqnarray*} \label{PB_Li_Lj}
			\{L_{i}' , L_{j}' \}_{nc} &=&\varepsilon_{i j \kappa}\bigg\{ \sum_{\eta,\varrho,\upsilon = 1}^{3}\bigg[\dfrac{1}{2} \varepsilon_{\eta \varrho \upsilon} \bigg(D_{\eta \varrho}q'^{\upsilon}q'^{\kappa} + E_{\eta \varrho}p'_{\upsilon}p'_{\kappa}\bigg) \\
			&+& \varepsilon_{\eta \varrho \kappa}\bigg( F_{\eta \kappa}p'_{\varrho}q'^{\kappa} - F_{\kappa \eta }q'^{\varrho}p'_{\kappa} -  F_{\eta \varrho }q'^{\kappa}p'_{\kappa} \bigg) \bigg] - F_{\kappa\kappa}L_{\kappa}'\bigg\},
	\end{eqnarray*}  }
	where $ i, j, h = 1,2,3,$
	yields
	$\{L_{i}' , L_{j}' \}_{nc} =  \varepsilon_{i j h} F'_{hh}L'_{h},$ where  $\varepsilon_{i j h} F'_{hh}$ are the structure constants of the Lie algebra.
	
	$\hfill{\square}$

	Moreover, the following relations hold:
	{\small \begin{align}
		& \{A_{i}' , A_{j}' \}_{nc} = - 2m\varepsilon_{i j h}F'_{hh}  H'L'_{h} \label{Ai_Aj} \\
		& \{L_{i}' , A_{i}' \}_{nc} =  F'_{jh} (L'_{h}p_{j}' + L'_{j}p_{h}' )  \label{Li_Ai} \\
		& \{L_{i}' , A_{j}' \}_{nc} = \varepsilon_{i j h} \bigg(F'_{h h} A_{h}' - F'_{h j}\dfrac{mk}{Y}q'^{j}  \bigg) +   F'_{h j} L'_{i}p_{h}, \label{Li_Aj}
		\end{align} }
	where $F'_{ij} = - F_{ij}$ and $ i,j,h = 1,2,3.$
	
	Let us now  consider some noncommutative constant energy hypersurfaces $ \Pi_{c}$ defined as:    {\small\[\Pi_{c} := \{(q,p)\in \mathcal{T^{\ast}Q}|
		H(q,p) = c \}, \]} where $c$ is a constant.
	Then, following \cite{lig} and \cite{die}, as the $1$-form $dH'$ obtained in \eqref{Eq_dh}  has no zeroes on $\mathcal{T^{\ast}Q},$  the  noncommutative constant energy hypersurfaces   $\Pi_{c}$  are closed submanifolds of $\mathcal{T^{\ast}Q}.$ Moreover,  let us define the following open submanifolds of $\mathcal{T^{\ast}Q}:$ \quad
	${\small\Pi_{\tau} :=\displaystyle\bigcup_{c\gtrless 0} \Pi_{c}, \quad  \{\tau = -, +\},}$
	such that
	{\small\begin{align*} \mathcal{T^{\ast}Q}=\Pi_{-}\cup \Pi_{0} \cup \Pi_{+},
		\Pi_{-}=\{(q,p)\in \mathcal{T^{\ast}Q}|H(q,p) < 0 \},
		\Pi_{+}=\{(q,p)\in \mathcal{T^{\ast}Q}|H(q,p) > 0 \} \end{align*}}
	with  $\Pi_{0}-$   the common boundary of
	$\Pi_{-}$ and $\Pi_{+}.$
	
	On  $\Pi_{\tau},$ we introduce
	{\small\begin{equation*}
		L_{i}'^{\tau}:= L_{i}'|\Pi_{\tau}, \quad \tau= \pm , \  i = 1,2,3,
		\end{equation*}}
	and define a scaled Runge-Lenz-Pauli vector $\hat{\Gamma}'$ by  
	{\small\begin{equation*}\label{fix1}
		\hat{\Gamma}'= \begin{cases}
		\hat{\Gamma}'^{-} =\dfrac{1}{(-2mH')^{1/2}}A', \ \ if \ \ \tau = -
		\\
		\hat{\Gamma}'^{+} = \dfrac{1}{(2mH')^{1/2}}A', \ \ if  \ \ \tau = + ,
		\end{cases}
		\end{equation*}}
	where $H'$ is the Hamiltonian function given in \eqref{Ham}.
	Further, we get $\{ H', \hat{\Gamma}'^{+}_{i} \}_{nc} = \{ H', \hat{\Gamma}'^{-}_{i} \}_{nc}=0$ proving that $\hat{\Gamma}'^{+}_{i}$ and $\hat{\Gamma}'^{-}_{i}$ are also constants of motion on $\Pi_{+}$ and $\Pi_{-},$ respectively. We can then rewrite
	the relations \eqref{Ai_Aj}, \eqref{Li_Ai}, and \eqref{Li_Aj} as follows:
	{\small\begin{eqnarray*}
			\{\hat{\Gamma}'^{\tau}_{i} , \hat{\Gamma}'^{\tau}_{j} \}_{nc} &=& -  \tau \varepsilon_{i j h} F'_{hh}   L'^{\tau}_{h} \label{gai_gaj}   \\
			\{L_{i}'^{\tau} , \hat{\Gamma}'^{\tau}_{i} \}_{nc} &=&  (2m\tau H')^{-1/2}F'_{jh} (L'^{\tau}_{h}p_{j}' + L'^{\tau}_{j}p_{h}' ) \label{Li_gai} \\
			\{L_{i}'^{\tau} , \hat{\Gamma}'^{\tau}_{j} \}_{nc} &=& \varepsilon_{i j h} \bigg(F'_{h h}\hat{\Gamma}'^{\tau} _{h} - (2m\tau H')^{-1/2}F'_{h j}\dfrac{mk}{Y}q'^{j}  \bigg) +   (2m\tau H')^{- 1/2}F'_{h j} L'^{\tau}_{i}p_{h},\label{Li_gaj}
	\end{eqnarray*} }
	where   $F'_{ij} = - F_{ij}, \ i,j,h = 1,2,3, \ \mbox{and} \ \{\tau = -, +\}. $
	
	Setting for all \ $i, j, h = 1, 2, 3,$  $L'^{\tau}_{h}p_{j}' = - L'^{\tau}_{j}p_{h}' $ and $ \dfrac{m}{Y}q'^{j} = \varepsilon_{i j h} L'^{\tau}_{i}p_{h}',$   we arrive at
	a Lie algebra isomorphic to the Lie algebra  $so(4)$ for  $ \tau = -:$
	{\small \begin{eqnarray*}
			\{L'^{-}_{i} , L'^{-}_{j} \}_{nc} &=& \varepsilon_{i j h} F'_{hh}L'^{-}_{h}   \\
			\{\hat{\Gamma}'^{-}_{i} , \hat{\Gamma}'^{-}_{j} \}_{nc} &=&    \varepsilon_{i j h} F'_{hh}   L'^{-}_{h}   \\
			\{L'^{-}_{i} , \hat{\Gamma}'^{-}_{j} \}_{nc} &=& \varepsilon_{i j h} F'_{h h}\hat{\Gamma}'^{-} _{h},
	\end{eqnarray*}}
	with the associated generators  given by
	{\small \begin{eqnarray*}
			& & \Phi_{hj} = \sum_{i = 1}^{3} \varepsilon_{h j i} F'_{ii}L'^{-}_{i}, \quad h,j,i = 1,2,3.  \\
			& & \Phi_{h4} = - \Phi_{4h} =  F'_{hh}\hat{\Gamma}'^{-}_{h}, \ \Phi_{44} = 0, \quad h = 1,2,3,
	\end{eqnarray*}}
	and,
	for $ \tau = +,$ at   a Lie algebra isomorphic to the Lie algebra $so(1,3):$
	{\small \begin{eqnarray*}
			\{L'^{+}_{i} , L'^{+}_{j} \}_{nc} &=& \varepsilon_{i j h} F'_{hh}L'^{+}_{h}   \\
			\{\hat{\Gamma}'^{+}_{i} , \hat{\Gamma}'^{+}_{j} \}_{nc} &= &   - \varepsilon_{i j h} F'_{hh}   L'^{+}_{h}   \\
			\{L_{i}'^{+} , \hat{\Gamma}'^{+}_{j} \}_{nc} &=& \varepsilon_{i j h} F'_{h h}\hat{\Gamma}'^{+} _{h},
	\end{eqnarray*}}
	with the corresponding generators provided by
	{\small \begin{eqnarray*}
			& & \Psi_{hj} = \sum_{i = 1}^{3} \varepsilon_{h j i} F'_{ii}L'^{+}_{i}, \quad h,j,i = 1,2,3.  \\
			& & \Psi_{h4} =  \Psi_{4h} =  - F'_{hh}\hat{\Gamma}'^{+}_{h}, \ \Psi_{44} = 0, \quad h = 1,2,3.  \label{psi}
	\end{eqnarray*}}
	
	\section{Case of  action-angle coordinates}  \label{sec5}
	The idea of considering a Liouville-integrable Hamiltonian system in the action-angle coordinates leads to many interesting results elucidating general properties of Hamiltonian systems \cite{abr,smir}.
	
	Assuming  the conditions
	{\small $\displaystyle\sum_{i,j,k=1}^{3}\lambda_{ij}\lambda_{ik}q^{j}q^{k} = 0$}  and {\small$ \alpha_{ij} = 0,$}  the Hamiltonian function \eqref{Ham} takes the form
	{\small \begin{align}\label{Ham2}
		H' = \dfrac{1}{2m}\sum_{i=1}^{3}\bigg[p_{i}^{2} - \dfrac{1}{2}\sum_{j,k = 1}^{3}\varepsilon_{ijk}L_{k} +  \dfrac{1}{4}\sum_{j=1}^{3}(\lambda_{ij}q^{j})^{2}\bigg] -\dfrac{k}{r},
		\end{align}}
	where   {\small$L_{k} = q^{j}p_{i} - q^{i}p_{j}, (i,j,k = 1,2,3) $} denote the components of the angular momentum $L$ on the cotangent bundle $\mathcal{T}^{\ast}\mathcal{Q} = \mathcal{Q} \times \mathbb{R}^{3}.$ The term {\small$ \varpi = - \dfrac{1}{4m}\displaystyle\sum_{i,j,k = 1}^{3}\varepsilon_{ijk}L_{k} +  \dfrac{1}{8m}\displaystyle\sum_{i,j=1}^{3}(\lambda_{ij}q^{j})^{2} $} contains the   deformation parameters perturbing the initial Kepler Hamiltonian function
	{\small\begin{align*}\label{Ham3}
		H = \dfrac{1}{2m}\sum_{i=1}^{3}p_{i}^{2}   -\dfrac{k}{r}.
		\end{align*}}
	Without loss of generality, let us consider the matrix  $\lambda$ under the form
	{\small\begin{equation*}
		\lambda =  \left(
		\begin{array}{cccc}
		0 & -\dot{\vartheta}\dot{\varphi}\sin(2\varphi) &\sqrt{2}\dot{\vartheta}\dot{\varphi}\cos{\varphi} \\
		\dot{\vartheta}\dot{\varphi}\sin(2\varphi)& 0&\sqrt{2}\dot{\vartheta}\dot{\varphi}\sin{\varphi} \\
		-\sqrt{2}\dot{\vartheta}\dot{\varphi}\cos{\varphi} & -\sqrt{2}\dot{\vartheta}\dot{\varphi}\sin{\varphi}& 0
		\end{array}
		\right),
		\end{equation*} }
	where we have $\varphi \in (0, 2\pi)$, $\vartheta \in (0, \pi),$ $\dot{\vartheta}\sin(2\varphi) << m, $   and $ \dot{\varphi}, \dot{\vartheta}$     are constants.
	There results the Hamiltonian function \eqref{Ham2}  in spherical-polar coordinates as follows:
	{\small\begin{align*}
		H' = \dfrac{1}{2m}\bigg[p_{r}^{2} + M^{2}\dfrac{p_{\vartheta}^{2}}{r^{2}} + \bigg( 1 + \dfrac{\dot{\vartheta}}{m}\sin(2\varphi)\bigg)\dfrac{p_{\varphi}^{2}}{r^{2}\sin^{2}{\vartheta}} \bigg] - \dfrac{k}{r},
		\end{align*}}
	where  
	\[ M =  \bigg(1 +  \dfrac{\sqrt{2}}{m}\dot{\varphi} + \dfrac{1}{2m}\dot{\varphi}^{2} \bigg)^{1/2}\quad \mbox{is a constant of motion,}  \]
	\[ p_{r} = m\dot{r},\ \  p_{\vartheta} = mr^{2}\dot{\vartheta}, \ \ \mbox{and} \ \ p_{\varphi} = mr^{2}\dot{\varphi}\sin^{2}\vartheta.
	\]
	The associated Hamiltonian vector field $ X_{H'} \in \mathcal{T}^{\ast}\mathcal{Q}$ corresponding  to the symplectic form:
	{\small\begin{equation*}
		\omega' = dp_{r}\wedge dr + dp_{\vartheta}\wedge d\vartheta + dp_{\varphi}\wedge d\varphi
		\end{equation*}}
	is given by
	{\small\begin{eqnarray*}
			X_{H'} & = & \dfrac{1}{m}\Bigg\{p_{r}\dfrac{\partial}{\partial{r}} -\dfrac{1}{r^{2}}\Bigg(mk - \dfrac{M^{2}p_{\vartheta}^{2}\sin^{2}\vartheta + \bigg( 1 + \dfrac{\dot{\vartheta}}{m}\sin(2\varphi)\bigg)p_{\varphi}^{2}}{r\sin^{2}\vartheta}\Bigg)\dfrac{\partial}{\partial{p_{r}}} +  M^{2}\dfrac{p_{\vartheta}}{r^{2}}\dfrac{\partial}{\partial{\vartheta}}
			\\
			& +& \bigg( 1 + \dfrac{\dot{\vartheta}}{m}\sin(2\varphi)\bigg)\dfrac{p_{\varphi}^{2}\cos\vartheta}{r^{2}\sin^{3}\vartheta}\dfrac{\partial}{\partial{p_{\vartheta}}}+ \bigg( 1 + \dfrac{\dot{\vartheta}}{m}\sin(2\varphi)\bigg)\dfrac{p_{\varphi}}{r^{2}\sin^{2}\vartheta}\dfrac{\partial}{\partial{\varphi}}   - \dfrac{\dot{\vartheta} \cos 2\varphi}{m r^{2} \sin^{2} \vartheta}\dfrac{\partial}{\partial p_{\varphi}}\Bigg\}.
	\end{eqnarray*}}
	Let us introduce  the following two additional integrals of motion:
	{\small\begin{equation} \label{integr}
		D_{\varphi} = \bigg( 1 + \dfrac{\dot{\vartheta}}{m}\sin(2\varphi)\bigg)^{1/2}p_{\varphi} , \  \mbox{and} \ \tilde{L}^{2} = M^{2} p_{\vartheta}^{2} + \dfrac{D_{\varphi}^{2}}{\sin^{2}\vartheta} ,
		\end{equation}  }
	where $\tilde{L}$ is the modified angular momentum vector in spherical-polar coordinates,  and
	{\small \begin{equation*} \label{cos}
		D_{\varphi}   = \tilde{L} \cos\xi.
		\end{equation*}}
	$\xi$ denotes the angle between the orbit plane and the equatorial plane $(x,y)$ \cite{vil2}.
	Since the Hamiltonian function $H'$ does not explicitly depend on the time, then setting $V = W - Et,$ it is possible to find an additive separable solution:
	{\small\begin{equation*}
		W = W_{r}(r)+   W_{\vartheta}(\vartheta) +  W_{\varphi}(\varphi),
		\end{equation*} }
	which reduces the Hamilton-Jacobi equation \cite{arn}
	{\small\begin{eqnarray*}
			\dfrac{\partial V}{\partial t} + H'\bigg(\dfrac{\partial V}{\partial q} \bigg/ q/t \bigg) = 0
	\end{eqnarray*}}
	to a simpler form
	{\small\begin{equation*}
		E= \dfrac{1}{2m}\bigg( \dfrac{\partial{W}}{\partial{r}}\bigg)^{2} + \dfrac{M^{2}}{r^{2}}\bigg( \dfrac{\partial{W}}{\partial{\vartheta}} \bigg)^{2} + \dfrac{1}{2mr^{2}\sin^{2}\vartheta}\bigg( 1 + \dfrac{\dot{\vartheta}}{m}\sin(2\varphi)\bigg)\bigg( \dfrac{\partial{W}}{\partial{\varphi}}\bigg)^{2} - \dfrac{k}{r},
		\end{equation*}}
	leading to the following set of equations:
	{\small \begin{equation*}
		\begin{cases}
		\bigg( \dfrac{dW_{\varphi} (\varphi)}{d\varphi}\bigg)^{2}= p_{\varphi}^{2} \\
		\bigg( \dfrac{dW_{\vartheta} (\vartheta)}{d\vartheta}\bigg)^{2} = \dfrac{1}{M^{2}} \bigg(\tilde{L}^{2}  - \dfrac{D_{\varphi}^{2}}{\sin^{2}\vartheta}\bigg)
		\\
		-r^{2}\bigg(\dfrac{dW_{r}(r)}{dr}\bigg)^{2} + 2mr^{2}E + 2mrk =
		\tilde{L}^{2}.
		\end{cases}
		\end{equation*}}
	In the compact case characterized by $E < 0,$  the action-angle variables  \cite{arn,liou,bog,smir} can be expressed as:
	{\small\begin{equation*}
		\begin{cases}
		J_{\varphi}= \dfrac{1}{2\pi}\oint \dfrac{dW_{\varphi}  (\varphi)}{d\varphi}d\varphi = \dfrac{1}{2\pi}\oint p_{\varphi} d\varphi = p_{\varphi}\\
		J_{\vartheta} = \dfrac{1}{2\pi } \oint  \dfrac{dW_{\vartheta}(\vartheta)}{d\vartheta}d\vartheta = \dfrac{1}{2\pi M} \oint \bigg(\tilde{L}^{2}  - \dfrac{D_{\varphi}^{2}}{\sin^{2}\vartheta}\bigg)^{1/2}d\vartheta  \\
		J_{r} =\dfrac{1}{2\pi} \oint \dfrac{dW_{r}(r)}{dr}dr = \dfrac{1}{2\pi} \oint \bigg( 2mE + \dfrac{2mk}{r} - \dfrac{\tilde{L}^{2}}{r^{2}} \bigg)^{1/2}dr
		\end{cases}
		\end{equation*} }
	{\small\begin{equation*} \label{angle}
		\varphi^{i} = \dfrac{\partial W}{\partial J_{i}}, \quad \varphi^{i}(0) = 0, \quad  i = 1, 2, 3.
		\end{equation*}}
	According to equation \eqref{integr}, $ J_{\varphi}$ can be re-written in terms of $D_{\varphi}$ and $\sin(2\varphi):$
	{\small\[J_{\varphi} =  D_{\varphi}\bigg( 1 + \dfrac{\dot{\vartheta}}{m}\sin(2\varphi)\bigg)^{-1/2}.\]}
	Since $\dot{\vartheta}\sin(2\varphi) << m, $ then $ \dfrac{\dot{\vartheta}\sin(2\varphi)}{m} \rightarrow 0$  and $\dfrac{\dot{\vartheta}\sin(2\varphi)}{2m} << 1.$ Using the first order Maclaurin expansion
	\cite{can},
	{\small\begin{equation*} \label{macl}
		\bigg( 1 + \dfrac{\dot{\vartheta}}{m}\sin(2\varphi)\bigg)^{-1/2} \backsimeq  1 - \dfrac{\dot{\vartheta}}{2m}\sin(2\varphi)  \backsimeq 1
		\end{equation*} }
	and the standard integration method \cite{born}, we get:
	{\small\begin{equation} \label{act-ang}
		\begin{cases}
		J_{\varphi} = J_{3}=D_{\varphi}\\
		J_{\vartheta} = J_{2} = \dfrac{1 }{ M} \bigg( \tilde{L} - D_{\varphi} \bigg) \\
		J_{r} = J_{1} = -\tilde{L}  + \dfrac{mk}{\sqrt{- 2mE}}\\
		\varphi^{1} = -\dfrac{1}{(J_{1} + MJ_{2} + J_{3})^{2}} \sqrt{\tilde{G}} + \arcsin\bigg[ \dfrac{mkr - (J_{1} + MJ_{2} + J_{3})^{2}}{\tilde{Q}}\bigg]\\
		\varphi^{2} = M \varphi^{1}  - M\arcsin\bigg[\frac{\bigg(1 - \dfrac{(MJ_{2}+ J_{3})}{mkr} \bigg)(J_{1} + MJ_{2} + J_{3})^{3/2} }{\tilde{Q}}\bigg] +  M \arcsin\bigg[\tilde{U}\cos\vartheta\bigg] \\
		\varphi^{3} = \dfrac{1}{M}  \varphi^{2} + \arcsin\bigg[\dfrac{J_{3}\cot\vartheta}{\sqrt{(MJ_{2}+ J_{3})^{2} - J_{3}^{2}}}  \bigg]  + \varphi,
		\end{cases}
		\end{equation} }
	where
	{\small\begin{align*}
		\tilde{Q} &= (J_{1} + MJ_{2} + J_{3})\sqrt{(J_{1} + MJ_{2} + J_{3})^{2} -(MJ_{2} + J_{3})^{2}}, \ \tilde{U} = \dfrac{(MJ_{2}+ J_{3})}{\sqrt{(MJ_{2}+ J_{3})^{2} - J_{3}^{2}}}, \cr
		\tilde{G} &= -m^{2}k^{2}r^{2} + 2mk(J_{1} + MJ_{2} + J_{3})^{2}r - (MJ_{2} + J_{3})^{2}(J_{1} + MJ_{2} + J_{3})^{2}.
		\end{align*} }
	Then, we obtain
	the Hamiltonian $H',$ the  Poisson bivector $P',$ the
	symplectic form $ \omega',$  and the Hamiltonian vector field $X_{H'}:$
	{\small\begin{eqnarray*} \label{Bil0}
			& & H' = E =  -\dfrac{mk^{2}}{2(J_{1} + MJ_{2} + J_{3} )^{2}},\ \  P'  = \sum_{h = 1}^{3}\dfrac{\partial}{\partial J_{h}} \wedge \dfrac{\partial}{\partial\varphi^{h}}, \\
			& & \omega'  =   \sum_{h = 1}^{3} dJ_{h} \wedge d\varphi^{h},\
			X_{H'} := \{H',.\} =
			\dfrac{mk^{2}}{(J_{1} + MJ_{2} + J_{3})^{3}}\bigg(
			\dfrac{\partial}{\partial{\varphi^{1}}} +
			M\dfrac{\partial}{\partial{\varphi^{2}}} + \dfrac{\partial}{\partial{\varphi^{3}}}\bigg),
	\end{eqnarray*}}
	satisfying the required relation  $\iota_{_{X_{H'}}} \omega'= -dH'.$   Therefore, in the action-angle coordinates $(J,\varphi),$  the triplet  $(\mathcal{T}^{\ast}\mathcal{Q},\omega',H')$ is also a Hamiltonian system.
	
	\section{ Construction of bi-Hamiltonian structures}\label{sec6}
	The generalized action-angle variables of Kepler's problem are usually denoted by $L,G,H, l = \mathtt{M},g= \omega,h = \Omega,$  called the Delaunay variables \cite{arn2}.
	$G$ and $H$ have the meaning of total and azimuthal angular momentum respectively, and $L$ has the meaning of a total orbital action. The angles $ \omega$, $\Omega $ are argument of periapsis and longitude of ascending node. The final angle $ \mathtt{M}$ is just the mean anomally containing the only real dynamics in the Kepler problem.
	
	Let us consider the following Delaunay-type variables:
	{\small\begin{equation*} \label{sys}
		\begin{cases}
		I_{1} &= J_{3} \\
		I_{2} &= MJ_{2} + J_{3} \\
		I_{3}&=  J_{1}+ MJ_{2} + J_{3}
		\end{cases};  \quad \begin{cases}
		\phi^{1} &= \varphi^{3} - \dfrac{1}{M}\varphi^{2}\\
		\phi^{2} &= \varphi^{2} - M\varphi^{1} \\
		\phi^{3} &= \varphi^{1},
		\end{cases}
		\end{equation*}}
	which coincide with the classical Delaunay variables for $ \lambda_{ij} =0$ \cite{morb}:
	{\small\begin{equation*} \label{dela}
		\begin{cases}
		I_{1}&\equiv H = \sqrt{mka(1 - \mathfrak{e}^{2})} \cos \xi \\
		I_{2}&\equiv G = \sqrt{mka(1 - \mathfrak{e}^{2})} \\
		I_{3}&\equiv L = \sqrt{mka}
		\end{cases};  \quad \begin{cases}
		\phi^{1}&\equiv h = \Omega \\
		\phi^{2}&\equiv g = \omega \\
		\phi^{3}&\equiv l = \mathtt{M} = n(t - t_{0}),
		\end{cases}
		\end{equation*}}
	where $\xi$ is the inclination, $n$ is the mean motion, $a$ is the semi-major axis of the orbit, $\mathfrak{e}$ is the eccentricity, and $t_{0}$ is the time at which the satellite passes through the perigee.
	
	Then, the  Hamiltonian function $H'$, the  symplectic form $\omega'$, the  Poisson bivector $P'$,  and the  Hamiltonian vector field $ X'_{H}$ are reduced to the expressions:
	{\small\begin{eqnarray*}
			H' =  - \dfrac{mk^{2}}{2 I_{3}^{2}}, \
			P'  = \sum_{j=1}^{3} \tilde{N}_{j}\dfrac{\partial}{\partial I_{j}} \wedge \dfrac{\partial}{\partial\phi^{j}}, \ \omega'  =  \sum_{j=1}^{3}\dfrac{1}{\tilde{N}_{j}} dI_{j} \wedge  d\phi^{j}, \
			X'_{H} = \dfrac{mk^{2}}{I_{3}^{3}}\dfrac{\partial}{\partial \phi^{3}},
	\end{eqnarray*}}
	where $\tilde{N }_{1} = 1,\tilde{N }_{2} = M, \tilde{N }_{3} = 1.$ 
	
	Recall that a vector field $X$ is called non-degenerate or anisochronous if the Kolmogorov condition \cite{bog} for the Hessian matrix
	{\small\begin{equation} \label{degn}
		det \bigg|  \frac{\partial^{2} H (J_{1},...,J_{n})}{\partial J_{i}\partial J_{k}}  \bigg|\neq 0
		\end{equation}}
	is met almost everywhere in the given action-angle coordinates. This condition implies that the dense subsets of the invariant $n$-dimensional tori of $X$ are closures of trajectories.
	If \eqref{degn} is not satisfied, $X$ is called degenerate or isochronous vector field.

	In our framework,  $X'_{H}$ is  a degenerate  Hamiltonian vector field since
	\begin{equation*} \label{degn_2}
	det \bigg|  \frac{\partial^{2} H'(J_{1},...,J_{n})}{\partial J_{i}\partial J_{k}}  \bigg| =  0.
	\end{equation*}
	Since our system is  isochronous  with a well-defined derivative for $H' = E < 0,$
	{\small\begin{equation*}
		\tilde{\mathfrak{a}} = \dfrac{\partial H'}{\partial J_{1}}= \dfrac{1}{k\sqrt{m}}(-2E)^{3/2},
		\end{equation*}}
	then according to the Bogoyavlenskij theorem \cite{bog}, we can get a bi-Hamiltonian formulation of this system in the domain of definition of the action-angle variables \eqref{act-ang}.  In particular, we use here the Hamiltonian function $H'$ in the defined Delaunay-type variable $(I,\phi)$  to construct  bi-Hamiltonian structures.
	Now, following  the example of the generic Bogoyavlenskij construction for the isochronous Hamiltonian system proposed by Grigoryev {\it et al.} in 2015 \cite{gri}, we can carry out the following canonical transformations:
	{\small\begin{align*}\label{cande}
		\tilde{I}_{1} = I_{1},\quad \tilde{I}_{2} = I_{2}, \quad \tilde{I}_{3} = H'= - \dfrac{mk^{2}}{2 I_{3}^{2}}, \quad \tilde{\phi}^{1} = \phi^{1}, \quad \tilde{\phi}^{2} = \phi^{2},
		\quad  \tilde{\phi}^{3} = \dfrac{k\sqrt{m}}{(-2H)^{3/2}} \phi^{3}.
		\end{align*}}
	permitting us to construct the following set of  Poisson bivectors for all $h \in \mathbb{N}$:
	{\small \begin{align*}
		\tilde{P}_{h} =  \tilde{\beta}_{1}(\tilde{I}_{1})\dfrac{\partial}{\partial\tilde{I}_{1}} \wedge \dfrac{\partial}{\partial\tilde{\phi}^{1}}  + \tilde{\beta}_{2}(\tilde{I}_{2})\dfrac{\partial}{\partial\tilde{I}_{2}} \wedge \dfrac{\partial}{\partial\tilde{\phi}^{2}}  +\bigg( \dfrac{dF_{h}}{d\tilde{I}_{3}}\bigg)^{-1}\dfrac{\partial}{\partial\tilde{I}_{3}} \wedge \dfrac{\partial}{\partial\tilde{\phi}^{3}}.
		\end{align*}}
	Putting
	{\small \begin{equation*}
		\tilde{\beta}_{1}(\tilde{I}_{1}) = \tilde{I}_{1}^{h} = I^{h}_{1} , \quad \tilde{\beta}_{2}(\tilde{I}_{2}) = M^{h+1} \tilde{I}_{2}^{h} = M^{h+1}I_{2}^{h}, \  \ F_{h} = - \dfrac{mk^{2}}{ (2 + h) I_{3}^{2 + h}}, \ h \in \mathbb{N},
		\end{equation*}}
	simplifies the expression of  the Poisson bivectors  $ \tilde{P}_{h}$ as
	{\small \begin{align*}
		\tilde{P}_{h} = \sum_{j = 1}^{3} \tilde{N}_{j}^{h + 1}I_{j}^{h} \dfrac{\partial}{\partial I_{j}} \wedge \dfrac{\partial}{\partial\phi^{j}}, \quad \tilde{N }_{1} = 1,\ \tilde{N }_{2} = M, \ \tilde{N }_{3} = 1.
		\end{align*}}
	Each  of $\tilde{P}_{h}$ is compatible with $P',$ $i.e.,$  $[\tilde{P}_{h}, P' ]_{NS} = 0.$ In this case, the eigenvalues of the corresponding recursion operator $T:= \tilde{P}\circ P'^{-1}$ are integrals of motion only \cite{bog,mar,smir2}.
	Then,  for all $h \in \mathbb{N},$ a hierarchy of Poisson bivectors $\tilde{P}_{h} $ and their corresponding $2$-forms $\tilde{\omega}_{h}$ are given by
	{\small\begin{equation*}
		\tilde{P}_{h} = \sum_{i,j = 1}^{6} (\tilde{P}_{h})^{ij}\dfrac{\partial}{\partial x^{i}}  \wedge \dfrac{\partial}{\partial x^{j}}, \quad \tilde{\omega}_{h} = \sum_{i,j = 1}^{6} (\tilde{\omega}_{h})_{ij} dx^{i}  \wedge  dx^{j},
		\end{equation*}}
	where $n =3, \ x^{k}=I_{k}, \ x^{k + 3} =
	\phi^{k}, \ k \leq 3,$
	{\small\begin{equation*}
		\begin{cases}
		(\tilde{P}_{h})^{14} &= I_{1}^{h} \\
		(\tilde{P}_{h})^{25} & = M^{h+1}I_{2}^{h}\\
		(\tilde{P}_{h})^{36} & = I_{3}^{h} \\
		(\tilde{P}_{h})^{ij} & = 0, \ \ \mbox{otherwise}
		\end{cases} , \qquad \begin{cases}
		(\tilde{\omega}_{h})_{41} &= I_{1}^{-h}\\
		(\tilde{\omega}_{h})_{52} &= M^{-(h+1)}I_{2}^{-h} \\
		(\tilde{\omega}_{h})_{63} &=  I_{3}^{-h} \\
		(\tilde{\omega}_{h})_{ij} &= 0 , \ \ \mbox{otherwise,}
		\end{cases},
		\end{equation*}
		$(\tilde{P}_{h})^{ij} = - (\tilde{P}_{h})^{ji}$,  and  $ (\tilde{\omega}_{h})_{ij} = -(\tilde{\omega}_{h})_{ji}.$}

	In the previous action-angle coordinate system $(J,\varphi),$ they  become
	{\small\begin{equation*}
		\begin{cases}
		(\tilde{P}_{h})^{14} &= \bigg(k\sqrt{\dfrac{-m}{2H'}} \bigg)^{h}  \\
		(\tilde{P}_{h})^{24} & = \dfrac{1}{M}\bigg((\tilde{P}_{h})^{14} - (\tilde{P}_{h})^{25} \bigg)\\
		(\tilde{P}_{h})^{25} & =M^{h}\tilde{L}^{h}  \\
		(\tilde{P}_{h})^{34} & = M(\tilde{P}_{h})^{24} \\
		(\tilde{P}_{h})^{35} &= M \bigg((\tilde{P}_{h})^{25}  - (\tilde{P}_{h})^{36}  \bigg)  \\
		(\tilde{P}_{h})^{36} & = J_{3}^{h} \\
		(\tilde{P}_{h})^{ij} & = 0, \ \ \mbox{otherwise}
		\end{cases} , \quad \begin{cases}
		(\tilde{\omega}_{h})_{41} &= \bigg(\dfrac{1}{k}\sqrt{\dfrac{-2H'}{m}} \bigg)^{h}  \\
		(\tilde{\omega}_{h})_{42} & = M\bigg((\tilde{\omega}_{h})_{41} - (\tilde{\omega}_{h})_{52} \bigg) \\
		(\tilde{\omega}_{h})_{52} & = \dfrac{1}{M^{h}\tilde{L}^{h}}  \\
		(\tilde{\omega}_{h})_{43} & = \dfrac{1}{M}(\tilde{\omega}_{h})_{42} \\
		(\tilde{\omega}_{h})_{53} &= \dfrac{1}{M} \bigg((\tilde{\omega}_{h})_{52}  - (\tilde{\omega}_{h})_{63}  \bigg) \\
		(\tilde{\omega}_{h})_{63} & = J_{3}^{-h} \\
		(\tilde{\omega}_{h})_{ij} & = 0, \ \ \mbox{otherwise,}
		\end{cases}
		\end{equation*}
		$(\tilde{P}_{h})^{ij} = - (\tilde{P}_{h})^{ji}$, $ (\tilde{\omega}_{h})_{ij} = -(\tilde{\omega}_{h})_{ji},$  $ H' = H'(J,\varphi),$ and $ \tilde{L}^{h} = \tilde{L}^{h}(J,\varphi).$}
	
	The  Poisson bracket $\{.,.\}_{\tilde{\omega}_{h}}$ with respect to each  symplectic form $\tilde{\omega}_{h}$  is now defined as:
	{\small\begin{equation*}
		\{ f,g\}_{\tilde{\omega}_{h}} = \sum_{i,j= 1}^{3} (\Lambda_{h})^{i}_{j}\bigg(\dfrac{\partial f}{\partial I_{i}} \dfrac{\partial g}{\partial \phi^{j}} -  \dfrac{\partial f}{\partial \phi^{j}} \dfrac{\partial g}{\partial I_{i}}\bigg), \  (\Lambda_{h}) = \left(
		\begin{array}{cccccc}
		(I_{1})^{h}& 0 & 0\\
		0& M^{h +1}I_{2}^{h} & 0\\
		0 & 0 & (I_{3})^{h} \\
		\end{array}
		\right).
		\end{equation*}}
	\begin{proposition}
		For each $h \in \mathbb{N},$ the vector field $X_{H}$ is a bi-Hamiltonian vector field  with respect to  $(\omega',\tilde{\omega}_{h}), $  {\it i.e.,}
		{\small\begin{equation*}
			\iota_{_{X_{H'}}} \omega' = - dH' \quad \mbox{and} \quad \iota_{_{X_{H'}}} \tilde{\omega}_{h} = - dF_{h}, \quad X_{H'} =  \{ H',.\} =  \{ F_{h} ,.\}_{\tilde{\omega}_{h}},
			\end{equation*}}
		where $F_{h}, (F_{0}\equiv H'),$ are integrals of motion for $X_{H'}.$
	\end{proposition}
	\textbf{Proof.}
	
	Since
	{\small\begin{equation*}
		\iota_{X}(df \wedge dg) = (Xf)dg - (df)Xg,
		\end{equation*}}
	we obtain
	{\small \begin{align*}
		\iota_{_{X_{H'}}} \omega'  = -dH',
		\quad
		\iota_{_{X_{H'}}} \tilde{\omega}_{h} = -dF_{h}, \quad \mbox{and} \quad X_{H'} =  \{ H',.\} =  \{ F_{h} ,.\}_{\tilde{\omega}_{h}}.
		\end{align*}}
	$\hfill{\square}$
	
	Besides, we obtain the recursion operators
	{\small\begin{equation*}
		T_{h} = \sum_{i,j = 1}^{3} (T_{h})_{i}^{j} \bigg(\dfrac{\partial}{\partial I_{j}}\otimes dI_{i}  + \dfrac{\partial}{\partial\phi^{j}} \otimes d\phi^{i} \bigg), \ h \in \mathbb{N}, \quad \begin{cases}
		(T_{h})^{1}_{1} &= I_{1}^{h} \\
		(T_{h})^{2}_{2} & = M^{h}I_{2}^{h}\\
		(T_{h})^{3}_{3} & = I_{3}^{h} \\
		(T_{h})^{i}_{j} & = 0, \ \ \mbox{otherwise}
		\end{cases}
		\end{equation*}}    
	taking   the following forms in      action-angle coordinate system $(J,\varphi):$
	{\small \begin{equation*}
		T_{h}= \sum_{i,j = 1}^{3}\bigg((R_{h})_{i}^{j}\dfrac{\partial}{\partial J_{j}} \otimes dJ_{i} + (S_{h})_{i}^{j}\dfrac{\partial}{\partial \varphi^{j}} \otimes d\varphi^{i}\bigg),
		\end{equation*}  }
	with
	{\small \begin{equation*}
		\begin{cases}
		(R_{h})^{1}_{1} &= (R_{h})^{3}_{3} = (R_{h})^{3}_{1} =  (\tilde{P}_{h})^{14}  \\
		(R_{h})^{2}_{1} &= M(\tilde{P}_{h})^{14} \\
		(R_{h})^{1}_{2} &= \dfrac{1}{M}(\tilde{P}_{h})^{14} \\
		(R_{h})^{2}_{2} &= (\tilde{P}_{h})^{14} + (\tilde{P}_{h})^{25} \\
		(R_{h})^{3}_{2} &= \dfrac{1}{M}(R_{h})^{2}_{2}\\
		(R_{h})^{2}_{3} &= M(R_{h})^{2}_{2} \\
		(R_{h})^{3}_{3} &=  (R_{h})^{2}_{2} + (\tilde{P}_{h})^{36}
		\end{cases} , \qquad \begin{cases}
		(S_{h})^{1}_{1} &= (R_{h})^{2}_{2}   \\
		(S_{h})^{2}_{1} &= - \dfrac{1}{M}(\tilde{P}_{h})^{25}\\
		(S_{h})^{1}_{2} &= - M(\tilde{P}_{h})^{25} \\
		(S_{h})^{2}_{2} &= (\tilde{P}_{h})^{36} + (\tilde{P}_{h})^{25}\\
		(S_{h})^{3}_{2} &= - M(\tilde{P}_{h})^{36} \\
		(S_{h})^{2}_{3} &= - \dfrac{1}{M}(\tilde{P}_{h})^{36} \\
		(S_{h})^{3}_{3} &= (\tilde{P}_{h})^{36}\\
		(S_{h})^{3}_{1} & = (S_{h})^{1}_{3} = 0.
		\end{cases}
		\end{equation*} }
	
	\section{Master symmetries}\label{sec7}
	In differential geometric terms, a vector field   $\Gamma$ on $\mathcal{T}^{\ast}\mathcal{Q}$ that satisfies
	{\small\begin{equation*}
		[X_{H'}, \Gamma] \neq 0 , \quad [X_{H'}, X] = 0, \quad [X_{H'}, \Gamma]  = X,
		\end{equation*}}
	is called a master symmetry or a generator of symmetries of degree  $m = 1$ for $X_{H'}$ \cite{cas,dam,fer2,ra,ra3}.
	
	In the following, we consider the Hamiltonian system $(\mathcal{T}^{\ast}\mathcal{Q}, \omega', H')$ and the integrals of motion $ F_{h},  h \in \mathbb{N}.$
	Thereby, we obtain the vector fields
	{\small\begin{equation*}
		X_{h} := \{F_{h},.\} = \dfrac{mk^{2}}{I_{3}^{(h + 3)}}\dfrac{\partial}{\partial \phi^{3}}
		\end{equation*}}
	which commute with the Hamiltonian vector field $X_{0} = X_{H'}$.
	The $X_{h}'s$ are called {\it dynamical symmetries} of the Hamiltonian system
	{\small $(\mathcal{T}^{\ast}\mathcal{Q}, \omega', H')$, $i.e., [X_{H'}, X_{h}] = 0.$}
	
	For the Hamiltonian system $(\mathcal{T}^{\ast}\mathcal{Q}, \omega', H')$ we introduce the  vector fields $\Gamma_{i\mu} \in \mathcal{T}^{\ast}\mathcal{Q}$
	{\small\begin{equation*}
		\Gamma_{i\mu} = \dfrac{1}{(3 + i)}  \sum_{j=1}^{3} \dfrac{\tilde{N}^{\mu}_{j}}{I_{j}^{(\mu - 1)}}\bigg(\dfrac{\partial}{\partial I_{j}} +  \dfrac{\partial}{\partial \phi^{j}}  \bigg), \ \    \tilde{N}_{1} = 1, \tilde{N}_{2} = M, \tilde{N}_{3} = 1, \ i,\mu \in \mathbb{N},
		\end{equation*}}
	satisfying the relation
	{\small\begin{equation*}
		\iota_{_{\Gamma_{i\mu}}}\omega' = -d\tilde{F}_{i\mu},
		\end{equation*}
		with
		\begin{equation*}
		\tilde{F}_{i\mu} =
		\begin{cases}
		\displaystyle\sum_{j=1}^{3} \dfrac{\tilde{N}_{j}}{3 + i} \bigg( \ln(I_{j}) - \dfrac{\phi^{j}}{I_{j}} \bigg), \quad \mu = 2 \\
		\displaystyle\sum_{j=1}^{3} \dfrac{\tilde{N}^{\mu - 1}_{j}}{3 + i} \bigg( \dfrac{I^{2 -\mu}_{j}}{2 - \mu} - \dfrac{\phi^{j}}{I^{\mu - 1}_{j}} \bigg), \quad \mu \neq 2.
		\end{cases}
		\end{equation*} }
	Computing the Lie bracket  between $X_{i}$ and $\Gamma_{i\mu},$ we obtain,  (see Fig $1$ for their diagram representation):
	{\small \begin{equation} \label{sym}
		[X_{i}, \Gamma_{i\mu}] = X_{i + \mu},\quad [X_{i}, X_{i + \mu}] = 0, \quad \mbox{with}\quad X_{i + \mu} = \dfrac{mk^{2}}{I_{3}^{(\mu + i +3)}}\dfrac{\partial}{\partial I_{3}}.
		\end{equation}}
	Hence, $\Gamma_{i\mu}$ are master symmetries or  generators of symmetries of degree  $m = 1$ for $X_{i}.$ The quantities $\tilde{F}_{i\mu}$ are called {\it master integrals}.
	\begin{figure}[!th]
		\centering
		\includegraphics[width=12cm, height=4cm]{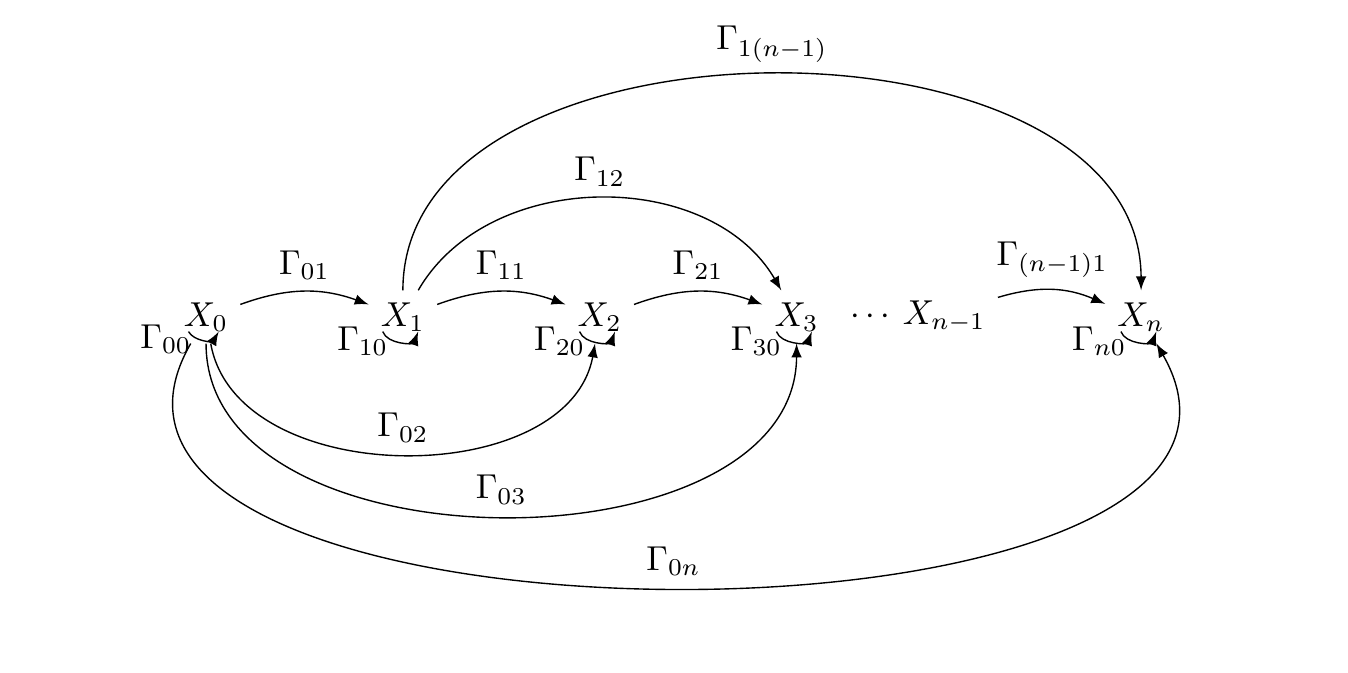}
		\caption{Diagrammatical illustration of equation \eqref{sym}.}
		\label{Graphs}
	\end{figure}
	Furthermore, we have
	{\small\begin{eqnarray*}
			& &\mathcal{L}_{\Gamma_{i0}} (P') = -\dfrac{1}{3 + i} P', \ (\tilde{\alpha} = -\dfrac{1}{3 + i}), \quad \mathcal{L}_{\Gamma_{i0}} (\tilde{P}_{1}) = 0, \ (\tilde{\beta} = 0), \cr
			& &  \mathcal{L}_{\Gamma_{i0}} (H') = -\dfrac{2}{3 + i} H',\ (\tilde{\gamma} = -\dfrac{2}{3 + i}),
	\end{eqnarray*}}
	showing that  the vector fields
	{\small\begin{equation*}
		\Gamma_{i0} = \dfrac{1}{(3 + i)}  \sum_{j=1}^{3} I_{j}\bigg(\dfrac{\partial}{\partial I_{j}} +  \dfrac{\partial}{\partial \phi^{j}}  \bigg)
		\end{equation*}}
	are conformal symmetries for both $P', \tilde{P}_{1}$ and $H'$ \cite{fer2}.
	Defining now the families of quantities $X'_{h}, \ \Gamma'_{ih}, \ P'_{h}, \ \omega'_{h} $ and $dH'_{h}$  by
	{\small\begin{equation*}
		X'_{h}:= T^{h}X_{0}, \ P'_{h}:= T^{h}P', \ \omega'_{h}:= (T^{\ast})^{h}\omega', \ \Gamma'_{ih}:= T^{h} \Gamma_{i0}, \ dH'_{h} :=(T^{\ast})^{h} dH',
		\end{equation*}}
	where $ i, h \in \mathbb{N}, $ and  $T^{\ast} :=  P'^{-1}\circ \tilde{P}_{1} $ denotes the adjoint of $T : = \tilde{P}_{1} \circ  P'^{-1}, $
	we obtain
	{\small \begin{align*}
		& P'_{h} = \sum_{j=1}^{3} \tilde{N}^{h+1}_{j} I^{h}_{j} \dfrac{\partial}{\partial I_{j}} \wedge \dfrac{\partial}{\partial\phi^{j}}, \  \Gamma'_{ih} =  \sum_{j =1}^{3} \dfrac{\tilde{N}^{h}_{j}I^{h + 1}_{j}}{3 + i}  \bigg(\dfrac{\partial}{\partial I_{j}} +  \dfrac{\partial}{\partial \phi^{j}}\bigg), \  X'_{h} = \dfrac{mk^{2}}{I_{3}^{3-h}}\dfrac{\partial}{\partial\phi^{3}}, \cr
		& \omega'_{h} = \sum_{j = 1}^{3} \tilde{N}^{h-1}_{j} I^{h}_{j} dI_{j}\wedge d\phi^{j} ,\ dH'_{h} = \dfrac{mk^{2}}{I_{3}^{3-h}}dI_{3} \ \mbox{and}  \ H'_{h} = \begin{cases}
		mk^{2} \ln(I_{3}),  \ \mbox{if} \ h =2,\\
		\dfrac{mk^{2}}{ (2-h)I_{3}^{2-h}}, \ \mbox{if} \ h \neq 2 \end{cases}
		\end{align*}}
	and, for each $i \in \mathbb{N},$ we derive the following plethora of conserved quantities:
	{\small\begin{align*}
		&\mathcal{L}_{\Gamma'_{ih}} (\Gamma'_{il}) = \dfrac{l - h}{3 + i}\Gamma'_{i(l +h)}, \ \mathcal{L}_{\Gamma'_{ih}} (X'_{l}) = - \dfrac{3 - l}{3 + i}X'_{l +h}, \ \mathcal{L}_{\Gamma'_{ih}} (P'_{l}) =  \dfrac{l - h -1}{3 + i}P'_{l + h},  \cr
		& \mathcal{L}_{\Gamma'_{ih}} (\omega'_{l}) = \dfrac{l + h + 1}{3 + i} \omega'_{l + h}, \ \mathcal{L}_{\Gamma'_{ih}} (T) =  \dfrac{1}{3 + i} T^{1 + h}, \quad (l \in \mathbb{N}), \cr
		& \langle dH'_{l} , \Gamma'_{ih} \rangle  = \begin{cases}
		\dfrac{mk^{2}}{3 + i},  \quad \mbox{if} \quad (h + l) =2,\\
		- \dfrac{2 - (h +l)}{3 + i} H'_{l + h}, \ \mbox{if} \ (h + l) \neq 2, \end{cases}
		\end{align*}}
	satisfying
	{\small\begin{align*}
		&\mathcal{L}_{\Gamma'_{ih}} (\Gamma'_{il}) = (\tilde{\beta} - \tilde{\alpha})(l - h)\Gamma'_{i(l +h)}, \  \mathcal{L}_{\Gamma'_{ih}} (X'_{l}) = (\tilde{\beta} + \tilde{\gamma} + (l - 1)(\tilde{\gamma} -\tilde{\alpha})) X'_{l + h}, \cr &\mathcal{L}_{\Gamma'_{ih}} (P'_{l}) = (\tilde{\beta} + (l -h -1)(\tilde{\beta} - \tilde{\alpha})) P'_{l + h}, \ \mathcal{L}_{\Gamma'_{ih}}(\omega'_{l})=  (\tilde{\beta} + (l + h + 1)(\tilde{\beta} - \tilde{\alpha})) \omega'_{l + h}, \cr
		& \mathcal{L}_{\Gamma'_{ih}}(T) = (\tilde{\beta} - \tilde{\alpha}) T^{1 + h} , \ \langle dH'_{l} , \Gamma'_{ih} \rangle  = \begin{cases}
		\dfrac{mk^{2}}{3 + i},  \ \mbox{if} \ (h + l) =2,\\
		(\tilde{\gamma} + (l + h)(\tilde{\beta} - \tilde{\alpha}))H'_{l + h}, \ \mbox{if} \ (h + l) \neq 2, \end{cases}
		\end{align*}}
	analogue to the  Oevel formulae (see \cite{oev,fer2,smir,smir2}).
	
	\section{Concluding remarks} \label{sec8}
	In this paper, we have defined a noncommutative phase space, derived
	a Hamiltonian system and proved the existence of dynamical symmetry
	groups $SO(3),$ $SO(4),$ and $SO(1,3)$ for the Kepler problem.
	Further, we have investigated the same Kepler problem in
	action-angle  coordinates and obtained its corresponding Hamiltonian
	system. Then, we have constructed a hierarchy of bi-Hamiltonian
	structures in the considered action-angle  coordinates following the
	example of the generic Bogoyavlenskij construction for the
	isochronous Hamiltonian system proposed by Grigoryev {\it et
		al.}, and   computed conserved quantities using  related master
	symmetries.
	
	\section*{Conflict of Interest}
	The authors declare that they have
	no conflicts of interest.
	
	\subsection*{Acknowledgments}
	The ICMPA-UNESCO Chair is in partnership
	with the Association pour la Promotion Scientifique de l'Afrique
	(APSA), France, and Daniel Iagolnitzer Foundation (DIF), France,
	supporting the development of mathematical physics in Africa.
	M. M. is supported by the Faculty of Mechanical Engineering,
	University of Ni\v s, Serbia, Grant ``Research and development of
	new generation machine systems in the function of the technological
	development of Serbia''.


\begin{thebibliography}{}
		\bibitem{abr}
		Abraham    R.,  Marsden J. E., \emph{Foundation of Mechanics ($2^{nd}$edition),} Addison-Wesley, New York (1978).
		\bibitem{arn}
		Arnold V. I.,\emph{ Mathematical Methods of Classical Mechanics}, Graduate Texts in Mechanics, \textbf{ 60}, Springer, New York (1978).
		\bibitem{arn2}
		Arnold V. I., Kozlov  V. V.,  Neishtadt A. I.,\emph{ Mathematical Aspects of Classical and Celestial Mechanics ($3^{rd}$ edition),}  Springer, \textbf{3}, (2006).
		\bibitem{bac}
		Bacry  H.,  Ruegg H., Souriau J. M.,
		\emph{Dynamical Groups and Spherical Potentials in Classical Mechanics}, Commun. Math. Phys., \textbf{3},  323-333 (1966).
		\bibitem{ban1}
		Bander M.,  Itzykson C., \emph{Group Theory and the Hydrogen Atom (I)}, Rev. Mod. Phys., \textbf{38},  330-345 (1966).
		\bibitem{ban2}
		Bander M.,  Itzykson C., \emph{Group Theory and the Hydrogen Atom (II)}, Rev. Mod. Phys., \textbf{38},  346-358 (1966).
		\bibitem{bar}
		Bargmann V., \emph{Zur Theorie des Wasserstoffatoms. Bemerkungen zur gleichnamigen Arbeit von V. Fock },  Z. Phys., \textbf{99},  576-582 (1936).
		\bibitem{bog}
		Bogoyavlenskij O. I., \emph{Theory of tensor invariants of integrable Hamiltonian systems. I. incompatible Poisson structures,} Commun. Math. Phys., \textbf{180},  529-586 (1996).
		\bibitem{born}
		Born M., \emph{The Mechanics of the Atom,}
		G.Bell And Sons Limited, London (1927).
		\bibitem{Brac}  Brackenridge J. B., \textit{The Key to Newton's Dynamics}, University of California Press (1995).
		\bibitem{can}
		Canute  C.,  Tabacco A., \textit{Mathematical analysis I,} Springer-Verlag, Milan (2008).
		\bibitem{cas}
		Caseiro  R., \textit{Master integrals, superintegrability and quadratic algebras,} Bull. Sci. Math., \textbf{126},  617-630 (2002).
		\bibitem{chan}
		Chang  D. E.,  Marsden J. E., \emph{Geometric Derivation of Delaunay Variables and Geometric Phases},  Celest. Mech. Dyn. Astron., \textbf{86},  185-208 (2003).
		\bibitem{chen}
		Chenciner  A., Montgomery  R., \emph{On a Remarkable Periodic Orbit of the Three-body Problem in the Case of Equal Masses},  Ann. Math., \textbf{152},  881-901 (2000).
		\bibitem{alinc2}
		Connes A., \emph{Geometrie noncommutative.} Inter Editions, Paris (1990).
		\bibitem{cuch}
		Cushman R. H., Duistermaat  J. J., \emph{A Characterization of the Ligon-Schaaf Regularization Map}, Comm. Pure and Appl., \textbf{50},  773-787 (1997).
		\bibitem{dam}
		Damianou P. A., \emph{Symmetries of Toda equations,} J. Phys. A, \textbf{26},  3791-3796 (1993).
		\bibitem{die}
		Dieudonn\'{e}  J., \emph{El\'{e}ments d'analyse}, Tome 3, Gauthiers-Villars, Paris (1970).
		\bibitem{du}
		Dubrovin B., \emph{Bihamiltonian structures of PDEs and Frobenius manifolds,} Lectures at the ICTP Summer School \textquotedblleft Poisson Geometry\textquotedblright, Trieste (2005).
		\bibitem{fer2}
		Fernandes R. L., \emph{On the master symmetries and bi-Hamiltonian structure of the Toda lattice,} J. Phys. A Math. Gen., \textbf{26},  3797-3803 (1993).
		\bibitem{fern}
		Fernandes  R. L., \emph{Completely integrable bi-Hamiltonian systems,} J. Dyn. Differ. Equ., \textbf{6}, No.1,  53-69 (1994).
		\bibitem{fil1}
		De Filippo S.,  Marmo G.,  Salerno M.,  Vilasi G., \emph{A New Characterization of Completely Integrable Systems}, Nuovo Cimento B, \textbf{83},  97-112 (1984).
		\bibitem{foc}
		Fock  V., \emph{Zur Theorie des Wasserstoffatoms},  Z. Phys., \textbf{98},  145-154 (1935).
		\bibitem{fra}
		Fradkin D. M.,  \emph{Existence of the dynamic symmetries $O_{4}$ and $SU_{3}$ for all classical central potential problems}, Prog. Theor. Phys., \textbf{37},  798-812 (1967).
		\bibitem{gel}
		Gelfand  I. M., Dorfman I. Y., \textit{The Schouten Bracket and Hamiltonian Operators,} Funct. Anal. Appl., \textbf{14},
		71-74 (1980).
		\bibitem{gui}
		Guichardet  A., \emph{Histoire d'un vecteur tricentenaire},
		Gaz. Math., \textbf{117},  23-33 (2008).
		\bibitem{gior}
		Gy\"{o}rgyi  G., \emph{Kepler's equation, Fock variables, Bacry's generators and Dirac brackets},
		Nuovo Cimento, \textbf{53A},  717-735 (1968).
		\bibitem{gri}
		Grigoryev Y. A., Tsiganov  A. V., \emph{On bi-Hamiltonian formulation of the perturbed Kepler problem,} J. Phys. A: Math. Theor., \textbf{48}, 175206(7pp) (2015).
		
		\bibitem{hkn1}
		Hounkonnou    M. N.,  Landalidji M. J.,  Balo\"{i}tcha E., \emph{Recursion Operator in a Noncommutative Minkowski Phase Space,}
		Proceedings of XXXVI Workshop on Geometric Methods in Physics, Poland 2017, Trends in Mathematics,  83-93 (2019).
		\bibitem{hkn2}
		Hounkonnou M. N.,  Landalidji M. J., \emph{Hamiltonian dynamics for the Kepler problem in a deformed phase space},
		Proceedings of XXXVII Workshop on Geometric Methods in Physics, Poland 2018, Trends in Mathematics, 34-48 (2019).
		\bibitem{hul}
		Hulthen  L., \emph{\"{U}ber die quantenmechanische Herleitung der Balmerterme},  Z. Phys., \textbf{86}, 21-23 (1933).
		\bibitem{kepl1}
		Kepler J., \emph{New Astronomy}, ed. William H. Donahue, Cambridge University Press, New York (1992).
		\bibitem{kepl2}
		Kepler J.,  \emph{The Harmony of the World},         ed. trans. E. J. Aiton, A. M. Duncan \& J. V. Field, American Philosophical Society, New York (1997).
		\bibitem{khos1}
		Khosravi N., Jalalzadeh S., Sepangi  H. R., \emph{Non-commutative multi-dimensional cosmology}, J. High Energy Phys., \textbf{01},  134 (2006).
		\bibitem{lax}
		Lax P. D., \emph{Integrals of nonlinear equations of evolution and solitary ways}, Commun. Pure Appl. Math., \textbf{21}, 467-490 (1968).
		\bibitem{len}
		Lenz   W., \emph{\"{U}ber den Bewegungsverlauf und die Quantenzust\"{a}nde des gest\"{o}rten Keplerbewegung}, Z. Phys., \textbf{24},  197-207 (1924).
		
		\bibitem{lig}
		Ligon  T., Schaaf M., \emph{On the global symmetry of the classical Kepler problem}, Reports on Math. Phys., \textbf{9}, 281-300 (1976).
		\bibitem{liou}
		Liouville  J., \emph{Note sur l'integration des \'{e}quations diff\'{e}rentielles de la dynamique,} J. Math. Pure Appl., \textbf{20},  137-138 (1855).
		.
		\bibitem{lio}  Liouville R., \textit{Sur le mouvement d'un corps solide
			pesant suspendu par l'un de ses points,} Acta Math. \textbf{20},
		239--284  (1897).
		\bibitem{liv}
		Livio M., \emph{The Golden Ratio: The Story of Phi, the World's Most Astonishing Number}, Broadway Books, New York (2003).
		
		\bibitem{mag1}  Magri F., \textit{A simple model of the integrable
			Hamiltonian equation,} J. Math. Phys., \textbf{19},
		1156-62 (1978).
		\bibitem{ma}
		Malekolkalami  B.,  Atazadeh K., Vakili  B., \emph{Late time acceleration in a non-commutative model of    modified cosmology}, Phys. Lett.B, \textbf{739},  400-404 (2014).
		\bibitem{marl}
		Marle C. M., \emph{A Property of Conformally Hamiltonian Vector Fields; Application to the Kepler Problem}, AIMS Journals, \textbf{4},  181-206 (2012).
		\bibitem{mar}
		Marmo  G.,  Vilasi G.,\emph{ When Do Recursion Operators Generate New Conservation
			Laws?}, Phys. Lett. B, \textbf{277},  137-140 (1992).
		\bibitem{mil}
		Milnor J.,\emph{ On the Geometry of the Kepler Problem}, Amer. Math. Monthly., \textbf{90},  353-365 (1983).
		\bibitem{morb}
		Morbidelli A., \emph{Modern Celestial Mechanics:
			Aspects of Solar System Dynamics,} Taylor \& Francis, \textbf{5}, New York (2002).
		\bibitem{mos}
		Moser J., \emph{ Regularization of Kepler's Problem and the Averaging Method on a Manifold,} Commun. Pure Appl. Math., \textbf{Vol. XXIII}, 609-636 (1 970).
		\bibitem{oev}
		Oevel W., \emph{A Geometrical Approach to Integrable Systems Admitting Time Dependent Inwatiants,} in Proceedings of the Conference on Nonlinear Evolution Equations, Solitons
		and the Inverse Scattering Transform, M. Ablowitz, B. Fuchssteiner and M. Kruskal eds., Oberwolfach (1986).
		\bibitem{pa}
		Pauli  W., \emph{\"{U}ber das Wasserstoffspektrum vom Standpunkt der neuen Quantenmechanik},  Z. Phys.,\textbf{ 36}, 336-363 (1926) .
		\bibitem{pau}
		Poincar\'{e} H., \emph{Sur les quadratures m\'{e}caniques} Acta Math., \textbf{13}, 1 (1899).
		\bibitem{ra}
		Ra\~{n}ada  M. F., \emph{A system of $n = 3$ coupled oscillators with magnetic terms: symmetries and integrals of
			motion,} SIGMA, \textbf{1}, 004, 7 pages  (2005).
		\bibitem{ra3}
		Ra\~{n}ada M. F., \emph{Superintegrability of the Calogero-Moser system: constants of motion, master symmetries, and time-dependent symmetries,} J. Math. Phys., \textbf{40}, 236-247 (1999).
		\bibitem{rud}
		Rudolph  G.,  Schmidt M.,  \emph{Differential
			geometry and mathematical physics, Part I. manifolds, Lie group and
			Hamiltonian systems,} Springer, New York (2013).
		\bibitem{san}
		Santoprete M., \emph{On the relationship between two notions of compatibility for bi-Hamiltonian systems,} SIGMA, \textbf{11}, 089, 11 pages (2015).
		\bibitem{smir}
		Smirnov R. G., \emph{Magri-Morosi-Gel'fand-Dorfman's bi-Hamiltonian constructions in the action-angle
			variables}, J. Math. Phys., \textbf{38},  6444 (1997).
		\bibitem{smir2}
		Smirnov R. G.,  \emph{The action-angle coordinates revisited: bi-Hamiltonian systems,} Rep. Math.
		Phys., \textbf{44},  199-204 (1999).
		\bibitem{vak}
		Vakili B.,  Pedram P.,  Jalalzadeh S., \emph{Late time acceleration in a deformed phase space model of dilaton cosmology}, Phys. Lett. B, \textbf{687}, 119-123 (2010).
		\bibitem{vil1}
		Vilasi G.,\emph{On the Hamiltonian Structures of the Korteweg-de Vries and Sine-
			Gordon Theories}, Phys. Lett. B, \textbf{94},  195-198 (1980).
		\bibitem{vil2}
		Vilasi G., \emph{Hamiltonian Dynamics}, World Scientific Publishing Co. Pte. Ltd., Singapore (2001).
		\bibitem{Voe}
		Voelkel J. R.,  \emph{Johannes Kepler and the New Astronomy}, Oxford University Press,
		New York (1999).
		\bibitem{zho}
		Zhou J.,  \emph{On Geometry and Symmetry of Kepler Systems. I.}, ( arXiv:1708.05504v1 [math-ph]), (2017).
	\end{thebibliography}
\end{document}